\documentclass[aps,prc,twocolumn,superscriptaddress,nofootinbib,10pt,floatfix]{revtex4-1}
\usepackage{graphicx}
\pdfoutput=1

\newcommand{\ifproofpre}[2]{#2}

\usepackage{graphicx}
\usepackage{amsmath}
\usepackage{framed}
\usepackage{tabularx}
\usepackage{pifont}
\usepackage{hyperref}
\usepackage{txfonts}
\usepackage{isotope}
\usepackage{upgreek}
\usepackage[sort&compress]{natbib}

\newcommand{\SI}[2]{\ensuremath{{#1}\,{#2}}}
\newcommand{\si}[1]{\ensuremath{#1}}
\newcommand{\ang}[1]{\ensuremath{{#1}^{\circ}}}
\newcommand{\num}[1]{\ensuremath{#1}}

\newcommand{\BEtwo}{B(E2; 3/2^- \rightarrow 1/2^-)}

\newcommand{\MHz}{{\mathrm{MHz}}}
\newcommand{\keV}{{\mathrm{keV}}}
\newcommand{\MeV}{{\mathrm{MeV}}}
\newcommand{\fm}{{\mathrm{fm}}}
\newcommand{\um}{{\mathrm{\upmu m}}}
\newcommand{\mm}{{\mathrm{mm}}}
\newcommand{\cm}{{\mathrm{cm}}}
\newcommand{\us}{{\mathrm{\upmu s}}}
\newcommand{\uCi}{{\mathrm{\upmu Ci}}}
\newcommand{\uA}{{\mathrm{\upmu A}}}
\newcommand{\torr}{\mathrm{Torr}}
\newcommand{\tesla}{\mathrm{T}}
\newcommand{\nucmag}{\mu_N}
\newcommand{\Nmax}{{N_\mathrm{max}}}
\newcommand{\hw}{{\hbar\omega}}
\newcommand{\ntwolo}{N\textsuperscript{2}LO}
\newcommand{\nthreelo}{N\textsuperscript{3}LO}
\newcommand{\percent}{\%}

\begin{document}

\title{First Measurement of the $\BEtwo$ Transition Strength in \isotope[7]{Be}: Testing \textit{Ab Initio} Predictions for $A=7$ Nuclei}

\author{S.~L.~Henderson}
\author{T.~Ahn}
\email{Corresponding author: tan.ahn@nd.edu}
\affiliation{Department of Physics and the Joint Institute for Nuclear Astrophysics, University of Notre Dame, 225 Nieuwland Science Hall, Notre Dame, IN 46556}
\author{M.~A.~Caprio}
\author{P.~J.~Fasano}
\affiliation{Department of Physics, University of Notre Dame, 225 Nieuwland Science Hall, Notre Dame, IN 46556, USA}
\author{A.~Simon}
\author{W.~Tan}
\author{P.~O'Malley}
\author{J.~Allen}
\author{D.~W.~Bardayan}
\author{D.~Blankstein}
\author{B.~Frentz}
\author{M.~R.~Hall}
\author{J.~J.~Kolata}
\affiliation{Department of Physics and the Joint Institute for Nuclear Astrophysics, University of Notre Dame, 225 Nieuwland Science Hall, Notre Dame, IN 46556}
\author{A.~E.~McCoy}
\affiliation{Department of Physics, University of Notre Dame, 225 Nieuwland Science Hall, Notre Dame, IN 46556, USA}
\affiliation{TRIUMF, Vancouver, British Columbia V6T 2A3, Canada}

\author{S.~Moylan}
\author{C.~S.~Reingold}
\author{S.~Y.~Strauss}
\affiliation{Department of Physics and the Joint Institute for Nuclear Astrophysics, University of Notre Dame, 225 Nieuwland Science Hall, Notre Dame, IN 46556}
\author{R.~O.~Torres-Isea}
\affiliation{Department of Physics, Randall Lab, 450 Church Street, University of Michigan, Ann Arbor, MI 48109, USA}

\date{\today}

\begin{abstract}
Electromagnetic observables are able to give insight into collective and
emergent features in nuclei, including nuclear clustering. These observables
also provide strong constraints for \textit{ab initio} theory, but comparison of
these observables between theory and experiment can be difficult due to the lack
of convergence for relevant calculated values, such as $E2$ transition
strengths. By comparing the ratios of $E2$ transition strengths for mirror
transitions, we find that a wide range of \textit{ab initio} calculations give
robust and consistent predictions for this ratio. To experimentally test the
validity of these \textit{ab initio} predictions, we performed a Coulomb
excitation experiment to measure the $\BEtwo$ transition strength in
\isotope[7]{Be} for the first time. A $\BEtwo$ value of
$26(6)_{\mathrm{stat}}(3)_{\mathrm{syst}}\,e^2\fm^4$ was deduced from the
measured Coulomb excitation cross section. This result is used with the
experimentally known \isotope[7]{Li} $\BEtwo$ value to provide an experimental
ratio to compare with the \textit{ab initio} predictions. Our experimental value
is consistent with the theoretical ratios within $1 \sigma$ uncertainty, giving
experimental support for the value of these ratios. Further work in both theory
and experiment can give insight into the robustness of these ratios and their
physical meaning.
\end{abstract}

\maketitle

\section{Introduction\label{sec:intro}}

Electromagnetic observables are sensitive probes of nuclear structure and have
sometimes yielded surprising and important results. For example, in heavy
nuclei, the discovery of nuclear deformation \cite{bohr1998:v1} and later
superdeformation  \cite{Twin1986}, both major advances in our understanding of
nuclear structure, have come from detailed studies of electromagnetic transition
strengths. The importance of electromagnetic probes and observables extend
outside of low-energy nuclear physics. For example, the use of high-energy
electron scattering has led to the elucidation of the charge distribution of the
neutron \cite{Gao2003,Hyde-Wright2004} and  the discovery of the EMC effect
\cite{Aubert1983,Gomez1994}, and continues to play a role in solving the proton
radius puzzle \cite{Pohl2013}. In light nuclei, the magnitude of electromagnetic
transition strengths can point to the existence of cluster states, halo nuclei,
or changes in nuclear deformation. For example, clustering enhances the $E2$
transition strength, due to clustered states having large deformation.

In addition, electromagnetic observables can provide a stringent test of
\textit{ab initio} nuclear theory. For instance, several electromagnetic
transition strengths have been determined to high precision in $A=10$ nuclei
using lifetime and branching ratio measurements and then compared to \textit{ab
  initio} quantum Monte Carlo calculations~\cite{McCutchan2009, McCutchan2012a,
  Lister2014, Kuvin2017}.  It was found that the calculated $E2$ transition
strengths were sensitive to the three-body interaction used.  However,
regardless of the three-body interaction, the calculations could not
consistently reproduce the $E2$ transition strengths across the $A=10$ isobars,
raising questions about the sufficiency of the interactions
used~\cite{McCutchan2012a}.

\textit{Ab initio} nuclear theory attempts to predict the properties of nuclei
starting directly from the description of the nucleus in terms of nucleons and
their
interactions~\cite{navratil2000:12c-ab-initio,pieper2004:gfmc-a6-8,neff2004:cluster-fmd,hagen2007:coupled-cluster-benchmark,epelbaum2011:12c-lattice-hoyle,bacca2012:6he-hyperspherical,shimizu2012:mcsm}.
The ingredients which comprise this formulation of the problem are well-defined:
once it is assumed that the nucleus can be treated as a system of nucleons
described by the nonrelativistic Schr\"odinger equation, then the energies and
wave functions of the nuclear eigenstates depend only on the internucleon
interaction~\cite{epelbaum2009:nuclear-forces}, which is the input to the
\textit{ab initio} theory. However, the internucleon interaction is imperfectly
known. It can only be partially determined from nucleon-nucleon scattering data.
Modern chiral effective field theory ($\chi$EFT) techniques aim to resolve the
ambiguities in the interaction by obtaining a systematic series expansion, in
which only a handful of low-energy constants remain to be determined from other
experimental inputs (such as pion-nucleon scattering or bound-state properties
of the $A=2$ and $3$ few-body systems~\cite{entem2003:chiral-nn-potl}).
Precision tests of the \textit{ab initio} predictions will be crucial in
validating the resulting $\chi$EFT description of nuclei.

An experimental test of \textit{ab initio} predictions, at least in principle,
directly tests the validity of the \textit{ab initio} framework and the inputs
entering into the \textit{ab initio} picture of the nucleus. However, to get
from the \textit{ab initio} formulation of the nuclear problem to concrete
\textit{ab initio} predictions for nuclear observables, we must overcome the
formidable practical challenge of obtaining accurate numerical solutions to the
many-body Schr\"odinger equation for the $A$-body system of interacting
nucleons. While several approaches have been developed for solving the
\textit{ab initio} nuclear many-body problem, including quantum Monte Carlo
(QMC) methods~\cite{pieper2004:gfmc-a6-8,carlson2015:qmc-nuclear} and the
no-core shell model
(NCSM)~\cite{navratil2000:12c-ab-initio,navratil2009:ncsm,barrett2013:ncsm} and
its
extensions~\cite{roth2007:it-ncsm-40ca,quaglioni2009:ncsm-rgm,dytrych2013:su3ncsm,romeroredondo2016:6he-correlations,mccoy2018:spncci-busteni17-URL},
each method is constrained by available computational resources. Only truncated
calculations of finite numerical accuracy can be carried out. The computed
observables, such as electromagnetic transition strengths, must be obtained with
sufficient accuracy to allow for meaningful comparison with experiment.

Although $E2$ transition strengths are observables of special interest due to
their sensitivity to nuclear shapes and deformation, the $E2$ operator is also
sensitive to the large distance ``tails'' of the nuclear wave function. It is
therefore especially challenging to obtain numerically converged \textit{ab
  initio} calculations of $E2$
strengths~\cite{maris2013:ncsm-pshell,caprio2015:berotor-ijmpe}. Inadequate
convergence precludes meaningful comparison of the calculated $E2$ strengths
with experiment, at least on an \textit{individual} basis.

However, we find that the \textit{ratios} of calculated $E2$ strengths for pairs
of transitions can indeed be well-converged, allowing for direct and meaningful
comparison with experiment. This is particularly true where the transitions
being compared involve states for which the wave functions all have similar
convergence properties. Notably, the wave functions for isospin mirror states
are closely related, making the comparison of $E2$ transitions in mirror nuclei
a particularly favorable case for obtaining precision tests of \textit{ab
  initio} theory.

The $\BEtwo$ transitions in the $A=7$ mirror nuclei $\isotope[7]{Li}$ and
$\isotope[7]{Be}$ therefore provide a natural opportunity for testing \textit{ab
  initio} theory. While the $\BEtwo$ ground state $E2$ transition strength in
$\isotope[7]{Li}$ is known from a number of Coulomb excitation experiments
~\cite{npa2002:005-007}, the corresponding $E2$ transition strength in
$\isotope[7]{Be}$ has never been measured. Since the decay of the $1/2^-$
excited state to the $3/2^-$ ground state is predominantly $M1$ in character,
the known lifetime of the $1/2^-$ level~\cite{Bunbury1956,Paul1966} only
provides information on the $M1$ transition strength. In contrast, Coulomb
excitation provides a viable mechanism for obtaining the $E2$ strength.

To measure the $\BEtwo$ transition strength in $\isotope[7]{Be}$, we have
performed a Coulomb-excitation experiment using a radioactive beam of
$\isotope[7]{Be}$. The measurement of this transition strength provides a rare
test for the $E2$ predictions of a large range of \textit{ab initio}
calculations~\cite{pervin2007:qmc-matrix-elements-a6-7,Pastore2013,Heng2017,Dohet-Eraly2016,QuaglioniPC},
involving a variety of traditional and chiral internucleon
interactions~\cite{wiringa1995:nn-av18,pieper2001:3n-il2,entem2003:chiral-nn-potl,shirokov2007:nn-jisp16,pieper2008:3n-il7-fm50,shirokov2016:nn-daejeon16,binder2016:lenpic-chiral},
and including new NCSM calculations presented here.

\section{Experiment\label{sec:expt}}

\begin{figure}
  \centering
  \includegraphics[width=\columnwidth]{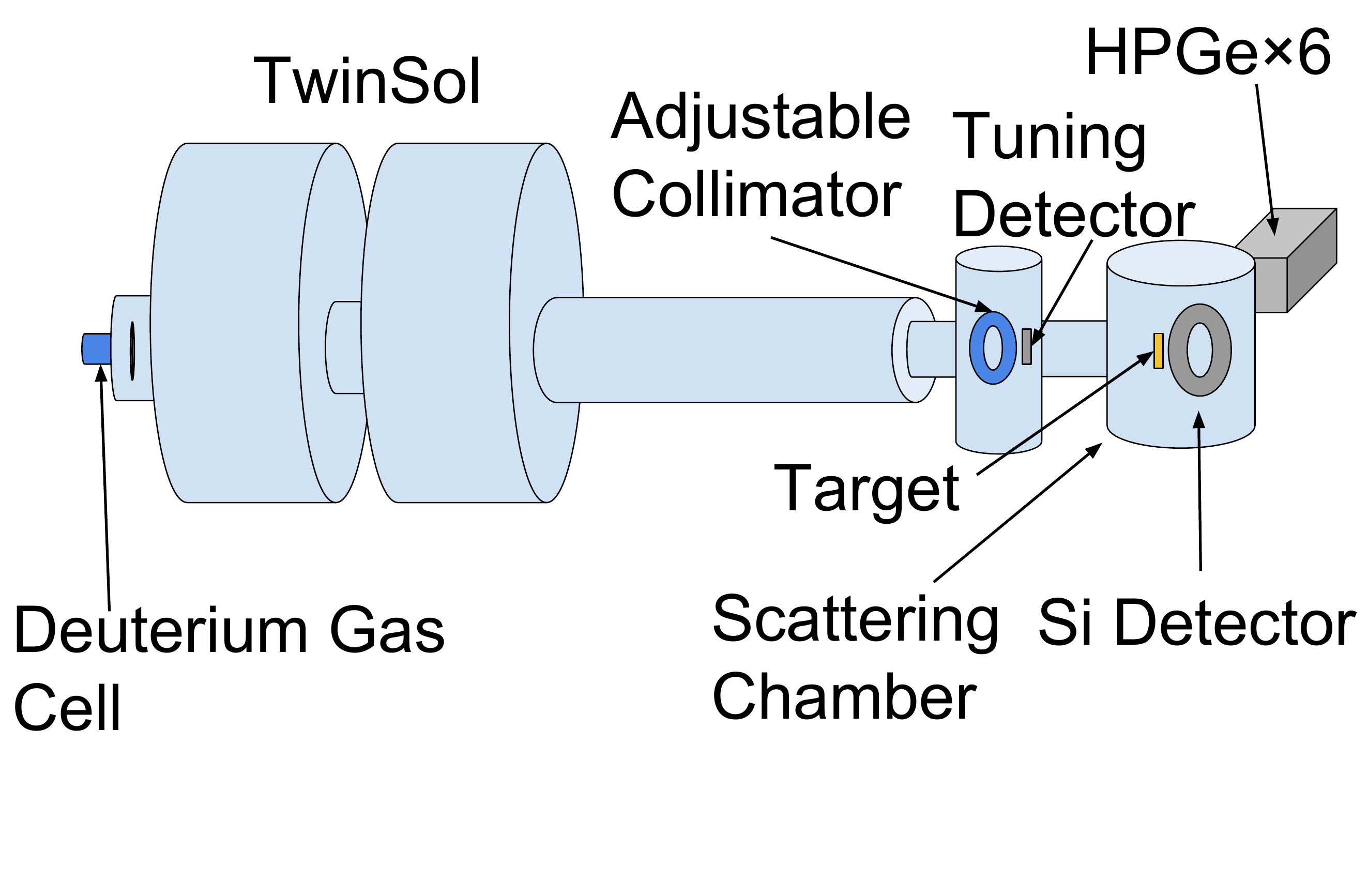}
  \caption{(Color online) A drawing showing the different components of the
  beamline, including Si and HPGe detectors, adjustable collimator, tuning
  detector, and Au target foil. Drawing is not to scale.}
  \label{fig:twinsol_beamline}
\end{figure}

\begin{figure}
  \centering
  \includegraphics[width=\columnwidth]{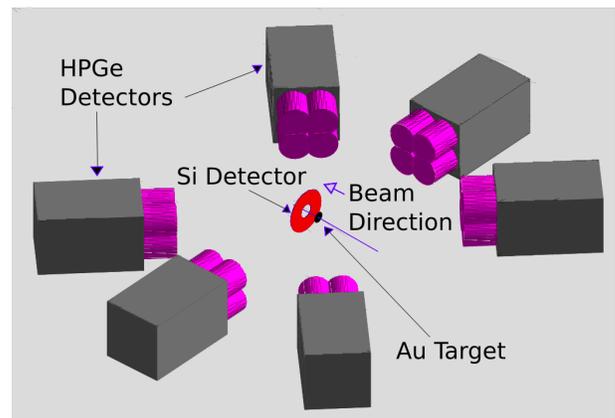}
  \caption{(Color online) The experimental setup is shown. Six HPGe Clover
  detectors placed at \ang{45}, \ang{90}, and \ang{135} with respect to the beam
  axis are shown surrounding the Au target and  S2 Si detector.}
  \label{fig:detector_setup}
\end{figure}

The Coulomb excitation experiment was performed using a radioactive beam of
\isotope[7]{Be} at the Nuclear Science Laboratory (NSL) located at the
University of Notre Dame. The NSL FN Tandem Van de Graaff accelerator was used
to accelerate a \SI{1.5}{e\uA} primary beam of \isotope[6]{Li} to
\SI{34.0}{\MeV}. By impinging the beam onto a \isotope[2]{H} gas cell at
\SI{800}{\torr}, we produced \isotope[7]{Be} through the
$\isotope[6]{Li}(\isotope[2]{H},n)\isotope[7]{Be}$ reaction. The secondary
\isotope[7]{Be} beam had an energy of \SI{31.3(10)}{\MeV} and was collected and
separated from competing reaction products using the two superconducting
solenoid magnets of \textit{TwinSol} \cite{Becchetti2003}. A diagram of the
\textit{TwinSol} beamline is shown in Fig.~\ref{fig:twinsol_beamline}. The first
solenoid was set at \SI{1.9}{\tesla} and the second at \SI{1.3}{\tesla} to
minimize the level of contaminants in the beam by focusing the beam through a
\SI{10}{\mm} diameter collimator at the crossover position between solenoids,
seen in Fig.\ref{fig:detector_setup}. More details on using \textit{TwinSol} for
$\gamma$-ray spectroscopy and Coulomb excitation can be found in
Refs.~\cite{Vincent2002,Amro2007,Brown1991}.

Downstream from \textit{TwinSol}, the beam was focused through an adjustable
collimator set to a \SI{9}{\mm} radius and then into the scattering chamber
\SI{35}{\cm} downstream from the collimator. The \isotope[7]{Be} beam was
initially tuned through the collimator onto a Si surface barrier detector on a
ladder directly after the collimator, then through an empty frame at the target
location. This Si tuning detector showed \SI{85}{\percent} of the beam to be
\isotope[7]{Be} with \isotope[6]{Li} and \isotope[7]{Li} comprising the majority
of the beam contaminants along with small amounts of \isotope[4]{He}. Both
contaminants had lower energies than the \isotope[7]{Be} ions and could be
separated in energy.

Inside the target chamber, the beam scattered off a \SI{1}{\um}-thick Au foil.
We selected Au for its high $Z$ and  the energy was chosen to be
\SI{77}{\percent} of the Coulomb barrier to eliminate any significant
contribution from the nuclear interaction. A \SI{300}{\um}-thick Micron
Semiconductor Limited S2 annular \isotope{Si} detector was placed \SI{25}{\mm}
downstream from the foil to measure the position and energy of the
\isotope[7]{Be} ions. A scaled drawing of the experimental setup in
Fig.~\ref{fig:detector_setup} shows the position of the \isotope{Si} detector
relative to the Au target foil. The Si detector has concentric ring electrodes
on the upstream side and radial sectors on the downstream side allowing the
measurement of particles scattering in an angular range of \ang{24}--\ang{55}.
The S2 Si detector rings begin \SI{11}{\mm} from the center of the detector and
end at \SI{35}{\mm} and there are 48 rings with \SI{0.5}{\mm} pitch. Pairs of
adjacent rings were electrically combined in the front-end feedthrough to make
24 rings, each effectively \SI{1}{\mm} wide each.

Outside of the scattering chamber, six  High-Purity Germanium (HPGe) clover
detectors from the Clovershare collaboration measured $\gamma$ rays in
coincidence with the \isotope[7]{Be} ions. The detectors were placed around the
gold foil, positioned \SI{20}{\cm} away and at \ang{45}, \ang{90}, and \ang{135}
from the beam axis. Bismuth Germanate (BGO) shields surrounded the HPGe
detectors. Although Compton suppression was not used in this experiment, the BGO
shields provided passive shielding from external background $\gamma$ rays and
the BGO shield hevimet collimators provided collimation for the $\gamma$ rays
produced in the experiment.

\begin{figure}
  \centering
  \includegraphics[width=\columnwidth]{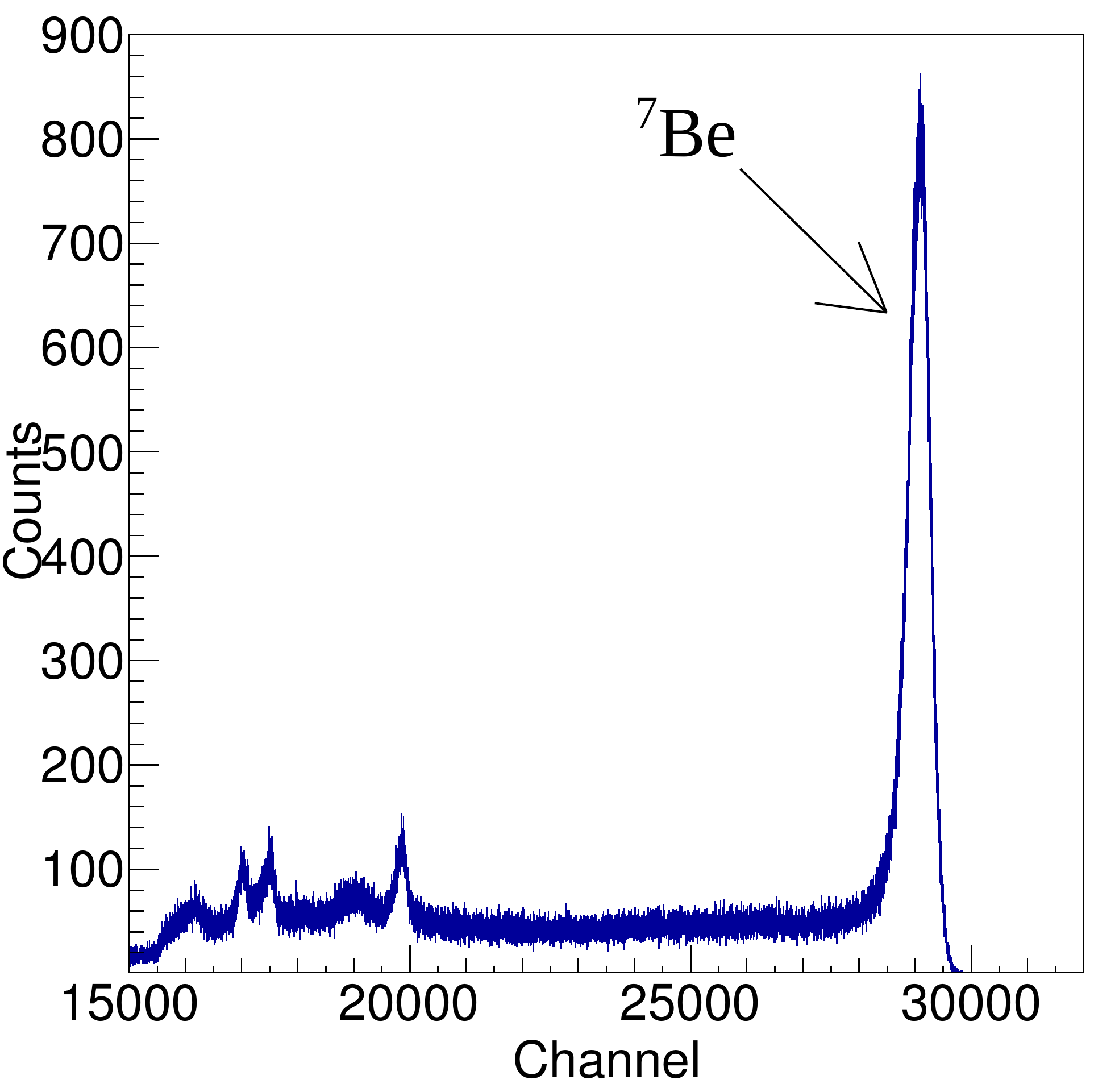}
  \caption{Shown above is an energy spectrum of particles seen in the 4th ring
  of our S2 particle detector from the center, corresponding to a nominal angle
  of \ang{31.8}. The data shown are only from ring events where a corresponding
  event in a sector of our detector was also seen. The high energy peak pictured
  is the elastically scattered $^7$Be while the smaller peaks are various
  contaminants of our beam that scattered through \textit{TwinSol} at lower
  energies.}
  \label{fig:ring_spectrum}
\end{figure}

Signals from both the Si and HPGe detectors were run through preamplifiers into
a digital data acquisition system, which had a sampling frequency of
\SI{100}{\MHz}. The data were written in list mode onto hard disk using the
Pixie-16 system \cite{Lipschutz2016}. An event was defined by a hit in a ring of
the Si detector with a coincidence timing window of \SI{2}{\us}, though only
events which saw hits in both a ring and a sector were used in the experiment.
An example spectrum of the different particles seen in the detector is shown in
Fig. \ref{fig:ring_spectrum}, with the central peak of the $^7$Be particles
separated from the lower energy contaminants.

The energy and efficiency of the HPGe detectors were calibrated with a
\SI{1.468}{\uCi} \isotope[152]{Eu} source. The detector array had a total
$\gamma$-ray efficiency of \SI{1.4}{\percent} at \SI{500}{\keV}. The energy
calibration was also verified by observing $\gamma$ rays from the Coulomb
excitation of \isotope[197]{Au} at their appropriate energies. The \SI{67}{\keV}
\isotope[197]{Au} $x$ ray and \SI{77}{\keV}, \SI{277}{\keV}, and \SI{547}{\keV}
$\gamma$ rays were seen. The energy resolution of our array was \SI{2.8}{\keV}
at \SI{1408}{\keV} and was sufficient for our measurement.

\section{Analysis}

The experimental analysis consisted of three major parts. First, the yield of
\isotope[7]{Be} $\gamma$ rays was determined from the Doppler-corrected spectrum
using recoil position information from the Si detector. Second, the integrated
beam flux was determined by comparing the measured rates of \isotope[7]{Be}
scatter to Monte Carlo simulations. Finally, by combining this information, the
$B({E}2)$ transition strength was calculated using the Winther-De Boer Coulomb
excitation code \cite{Winther1965}. The details of the analysis are presented
below.

\begin{figure}
  \centering
  \includegraphics[width=\columnwidth]{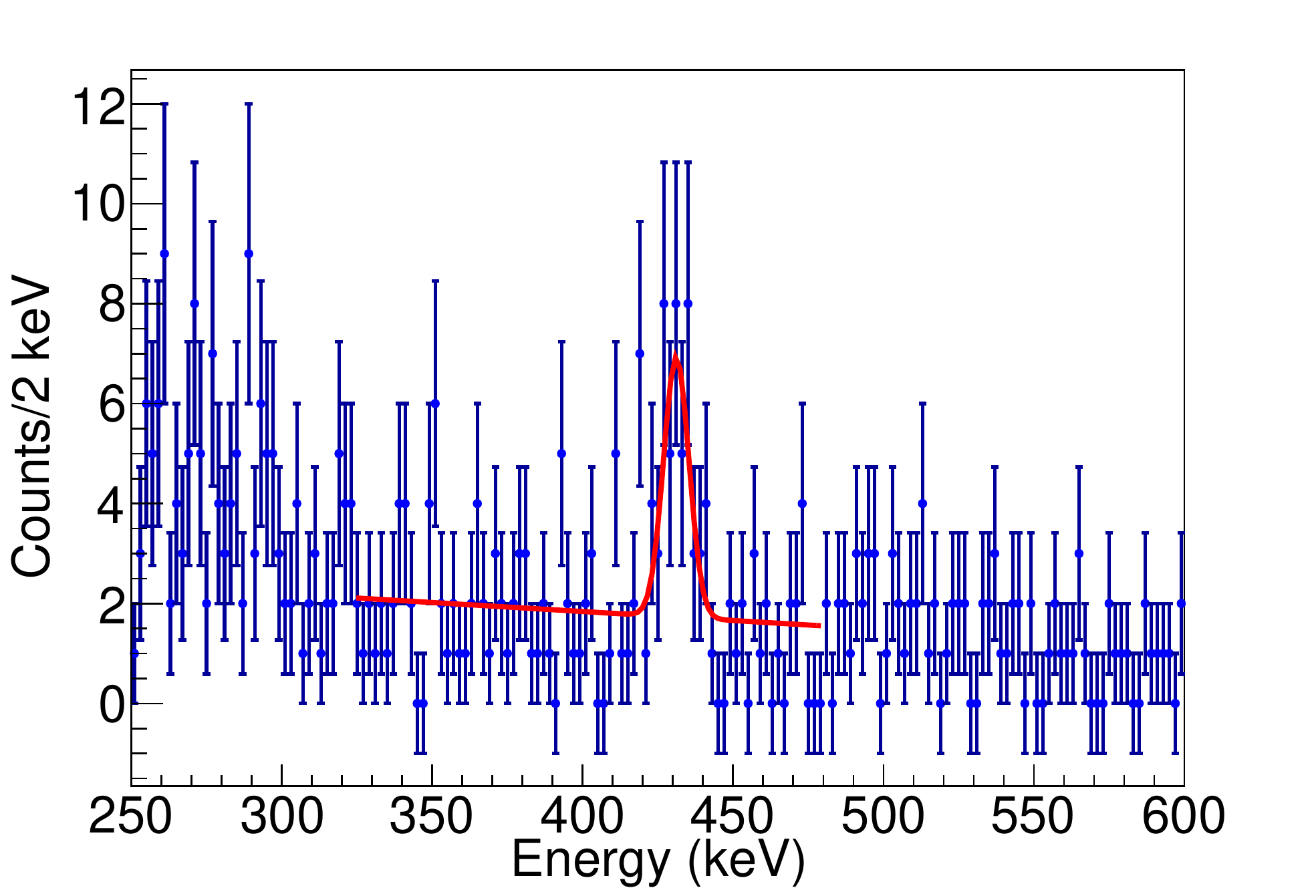}
  \caption{(Color online) The total, Doppler-corrected $\gamma$-ray spectrum,
  taken in coincidence with particles seen in the silicon detector, is shown.
  The spectrum is binned with 2 keV/bin. The $\gamma$-ray peak corresponding to
  the $1/2^- \rightarrow 3/2^-$ transition of \isotope[7]{Be} is seen at
  \SI{431}{\keV}.}
  \label{fig:gamma_spectrum}
\end{figure}

The direction of the \isotope[7]{Be} ions detected in the Si detector was used
to correct for the Doppler shift of the $\gamma$ rays emitted in flight. Random
coincidences were eliminated by requiring a tight time coincidence between the
Si and Ge detector signals. The Doppler-corrected spectrum yielded a peak with a
centroid value of \SI{431}{\keV} with a FWHM of \SI{10}{\keV}. This energy
corresponds to the $1/2^- \rightarrow 3/2^-$ transition of $^7$Be  and is shown
in Fig.~\ref{fig:gamma_spectrum}. We fit our $\gamma$-ray peak with a Gaussian
function and a linear background, which yielded a total peak area of
\SI{30(6)}{counts}. The calibrated efficiency of the HPGe array was used to
determine our final $\gamma$ yield.

\begin{figure}
  \centering
  \includegraphics[width=\columnwidth]{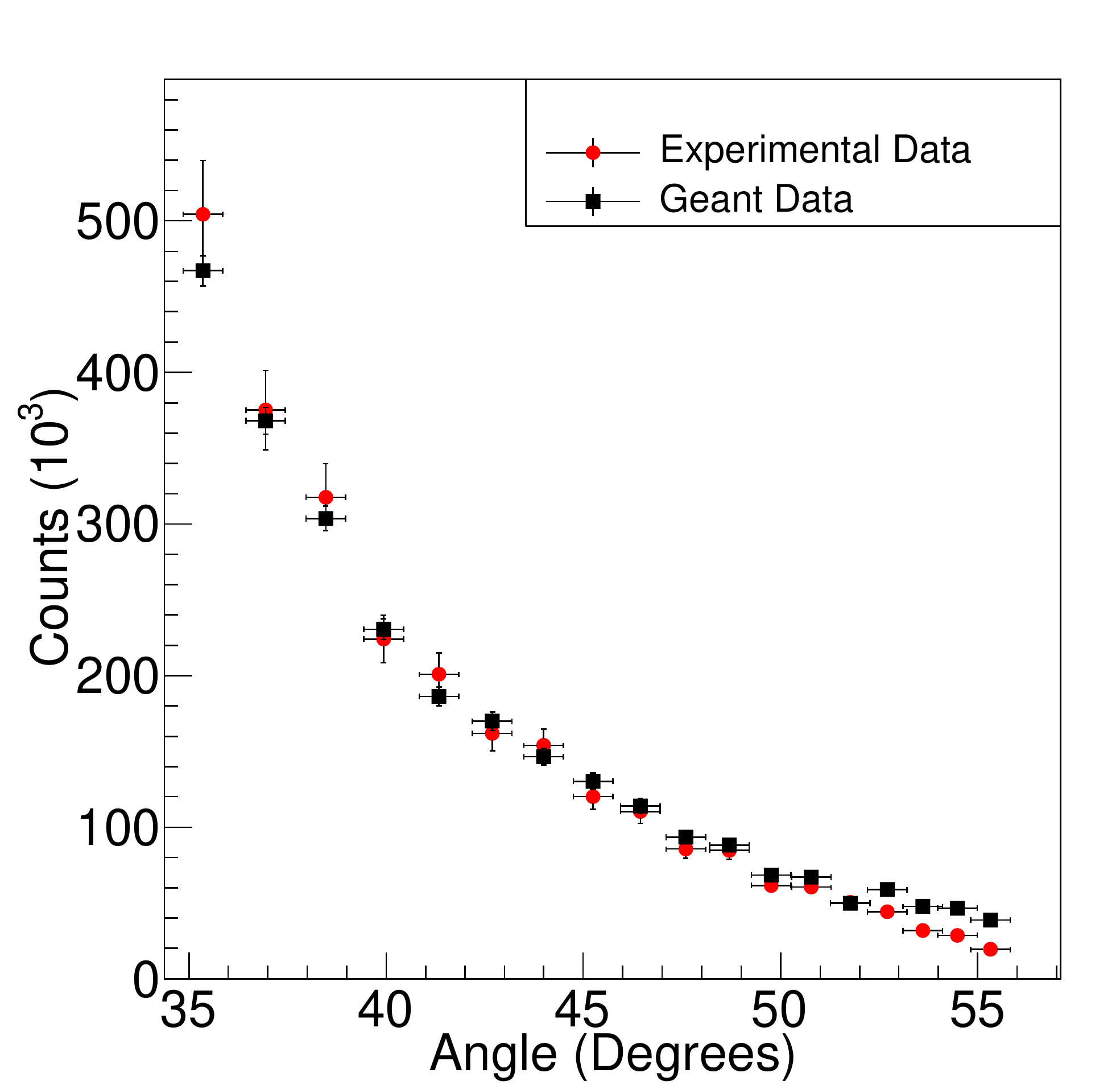}
  \caption{(Color online) A plot of \isotope[7]{Be} ions measured in the rings
  of the Si detector (circles) and the Geant4 simulated data (squares).}
  \label{fig:particleSpectrum}
\end{figure}

Determining the \isotope[7]{Be} beam flux on the \isotope{Au} target was a
necessary step in calculating the $B({E}2)$ value. Production of in-flight beams
with \textit{TwinSol} typically produces extended spot sizes on target. A LISE++
\cite{Tarasov2008} calculation of the beam transport through \textit{TwinSol} to
the Au target showed a fairly uniform beam with a radius on the order of
\SI{5}{\mm}. Due to the diffuseness of the beam and the proximity of the target
to the Si detector, the rings of the Si detector detected \isotope[7]{Be} ions
from a range of scattering angles. The Si detector sectors can also show
asymmetry in the measured rates if the incident beam is offset. To properly
account for these effects, a Geant4
\cite{Agostinelli2002,Allison2006,Allison2016} simulation was performed to
deduce the beam rate on target. Two beam parameters were varied in the
simulation: the beam radius and the offset from the beam axis. The angular
spread for the incident beam was also considered but we found it had little
impact on the simulation. These beam parameters were scanned over a range of
values (\SI{2}{\mm}--\SI{8}{\mm} for the radius and \SI{3}{\mm}--\SI{7}{\mm} for
the offset) to find the optimal parameters that reproduced the distribution seen
in the rings and sectors. A \SI{4}{\mm} beam radius and a \SI{4}{\mm} offset
best reproduced the shape of the Si detector ring and sector data. The
reproduction of the experimental data seen in the Si detector rings is shown in
Fig.~\ref{fig:particleSpectrum}. The agreement between the Geant4 simulation and
the data was good and yielded a beam rate of $8.8(4) \times 10^4$~pps. The
uncertainty in the beam rate was estimated by how much the beam parameters can
be changed before the shape of the beam exceeded the experimental uncertainties.

Next, the \isotope[7]{Be} $B({E}2)$ was calculated using a version of the
Winther-De Boer Coulomb excitation code  modified to perform calculations for
electric dipole to hexadecapole transitions based on the semi-classical theory
of Coulomb excitation \cite{Alder1956}. The Winther-De Boer code calculates
differential cross sections as a function of angle given an ${E}2$ matrix
element. The ${E}2$ matrix element was varied to reproduce the $\gamma$-ray
yield measured in the experiment. Because the \isotope[7]{Be} beam was broad,
the different scattering angles and detector geometric efficiencies were
accounted for using the Geant4 simulation mentioned above. The \isotope[7]{Be}
$\BEtwo$ value we obtained is \SI{26(6)}{e^2\fm^4}, which includes the
statistical and beam rate uncertainties. This value differs from our previous
reported preliminary result of \SI{34(8)}{e^2\fm^4} \cite{Ahn2018} due to the
use of a default value of the $E1$ dipole polarizability term in our previous
calculation. This default value was initially thought to be negligible but
turned out to significantly modify the Coulomb excitation cross section. We have
verified that we are able to completely turn off the contribution from the $E1$
polarizability term in our current calculation and have also checked our
calculated cross sections by using the same input parameters with the
coupled-channels code FRESCO \cite{fresco} and found the results consistent.

There were a number of systematic uncertainties associated with the measurement.
Two important considerations in deducing the $\BEtwo$ transition strength are
the influence of second-order processes and the contribution of the ${M}1$
excitation to the Coulomb excitation cross section. Of the second-order
processes, the largest is the virtual ${E}1$ excitation to the
\isotope[3]{He}-$\alpha$ breakup channel, the ${E}1$ dipole polarizability. This
effect is known to alter the Coulomb excitation cross section on the order of 10
percent at the energy and angles used on our experiment, based on the estimated
dipole polarizability seen in \isotope[7]{Li} \cite{Hausser1973}. This is due to
the low-energy threshold for breakup, which is at \SI{1.59}{\MeV}. The total
${M}1$ excitation contributing to the Coulomb excitation cross section is
calculated to be less than \SI{3}{\percent} for forward angles. Due to these
effects, we make a combined estimate for our systematic uncertainty as
\SI{13}{\percent} and \SI{3}{\percent} for the effect of the ${E}1$ dipole
polarizability and the ${M}1$ excitation, respectively. This gives a value of
\SI{\pm 3}{e^2\fm^4}. The uncertainties in our measurement stem primarily from
this systematic uncertainty due to ${E}1$ excitations and to the limited
statistics gathered in the experiment.

\section{Comparsion with \textit{Ab Initio} Theory}
\label{sec:theory}

\begin{figure*}
  \centering
  \includegraphics[width=\ifproofpre{0.85}{0.85}\hsize]{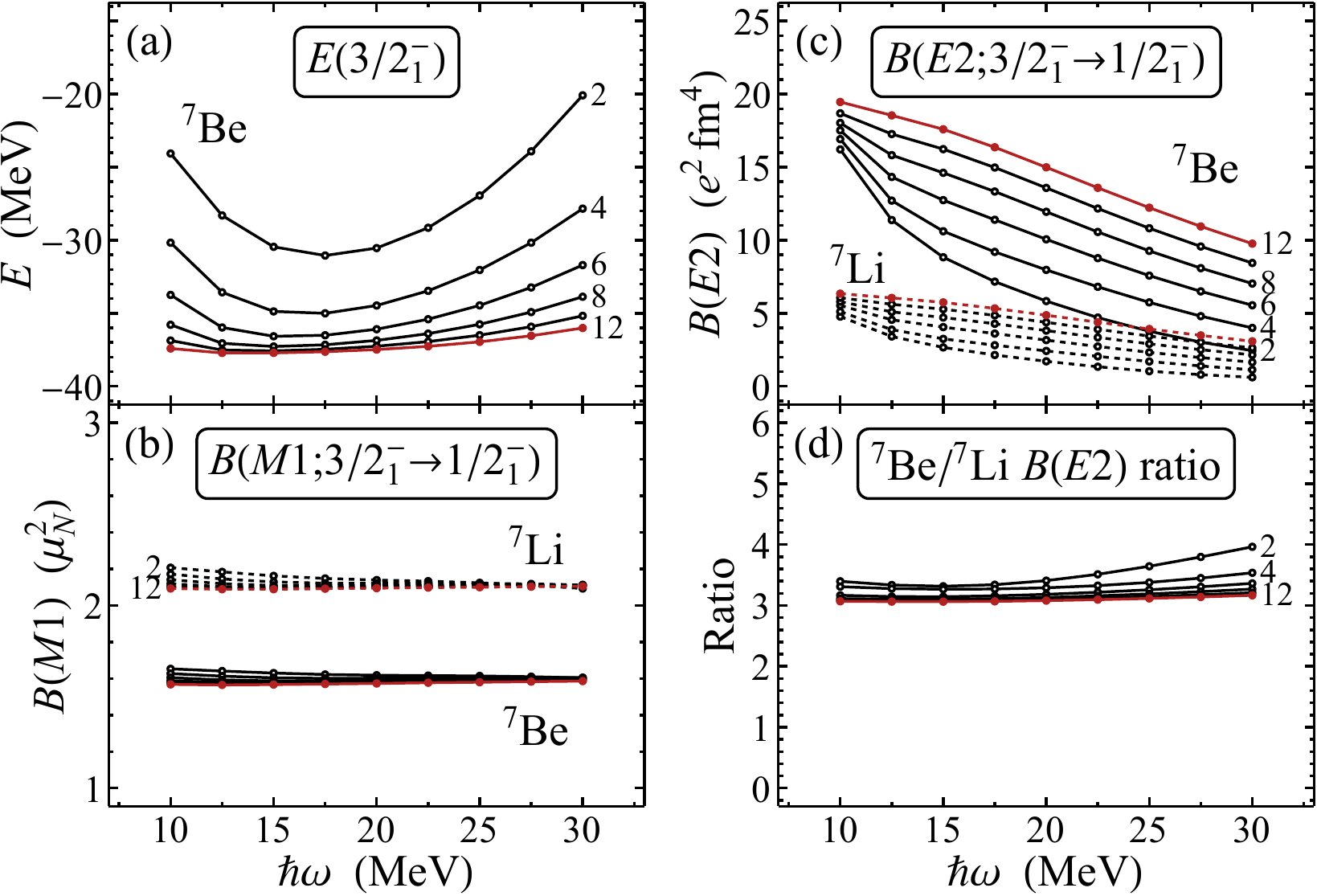}
  \caption{ Convergence of \textit{ab initio} NCSM calculations for
    $\isotope[7]{Li}$ and $\isotope[7]{Be}$, with the Daejeon16 interaction:
    (a)~the $3/2^-$ ground state energy ($\isotope[7]{Be}$ only), (b)~$\BEtwo$
    for $\isotope[7]{Li}$ (dashed curves) and $\isotope[7]{Be}$ (solid curves),
    (c)~$B(M1; 3/2^- \rightarrow 1/2^-)$ for $\isotope[7]{Li}$ (dashed curves)
    and $\isotope[7]{Be}$ (solid curves), and (d)~the ratio of $\BEtwo$
    strengths in $\isotope[7]{Li}$ and $\isotope[7]{Be}$.  Calculated values are
    shown as functions of the basis parameter $\hw$, for $\Nmax=2$ to $12$ (as
    labeled).  }
  \label{fig:convergence}
\end{figure*}

To use the present experimental result for the $\BEtwo$ strength in
$\isotope[7]{Be}$ as a test of \textit{ab initio} theory, we must contend with
the convergence limitations described in the introduction.  Recall that the
mathematical problem to be solved in the \textit{ab initio} nuclear description
of the nucleus is well-defined: find the eigenvalues and eigenfunctions of the
many-body Schr\"odinger equation, for $A$ nucleons, which are interacting by a
given internucleon interaction.  However, this is a formidable computational
problem, and the accuracy of the solutions is limited by available computational
power.  Observables which are sensitive to the long-range physics of the nucleus
(the tails of the nuclear wave function), such as $E2$ matrix elements and
charge radii, can be particularly challenging to compute.

Only when we have adequately addressed these numerical challenges can we compare
the results with experiment, and use this comparison as a meaningful test of the
predictive power of \textit{ab initio} nuclear theory.  Recall that the
fundamental input to the \textit{ab initio} description is the imperfectly-known
internucleon interaction entering into the many-body Schr\"odinger
equation.%
\footnote{\label{fn-convergence}Since nucleons are not simply point particles,
  electromagnetic observables calculated from the \textit{ab initio} wave
  functions also depend upon the electromagnetic current operators for the
  nucleons~\cite{Pastore2013}.  These current operators may need significant
  corrections from, \textit{e.g.}, meson-exchange currents, going beyond the
  single-nucleon impulse approximation.  Chiral approaches likewise provide a
  systematic approach to determining the current
  operators~\cite{park1996:chiral-vector-currents}.}

We start by noting that the Green's function Monte Carlo (GFMC) approach is able
to directly provide calculations of absolute $E2$ strengths, with well-defined
statistical uncertainties from the Monte Carlo calculation.  Predictions for the
absolute $B(E2)$ strengths in $\isotope[7]{Li}$ and $\isotope[7]{Be}$, from
Refs.~\cite{pervin2007:qmc-matrix-elements-a6-7,Pastore2013}, are shown in
Table~\ref{tab:be2}, along with the experimental values.  These calculations are
based on an internucleon interaction with an AV18 two-body
part~\cite{wiringa1995:nn-av18} and either an IL2~\cite{pieper2001:3n-il2} or
IL7~\cite{pieper2008:3n-il7-fm50} three-body contribution.  The calculated
values for the $E2$ strength in $\isotope[7]{Li}$ are generally consistent with
the measured value.  There is significantly greater variation among the
calculated values for the $E2$ strength in $\isotope[7]{Be}$, with these values
lying either just inside or just outside the lower edge of the uncertainty on
the present measured value.

However, the GFMC approach is limited in its ability to accommodate
state-of-the-art nonlocal chiral EFT
interactions~\cite{carlson2015:qmc-nuclear,piarulli2016:local-chiral-potentials}.
There can furthermore be systematic effects in the many-body
calculation~\cite{carlson2015:qmc-nuclear}, \textit{e.g.}, from
clusterization~\cite{McCutchan2012a}, which may not be accounted for in the
statistical uncertainties.  It is therefore important to also move forward with
comparisons against NCSM results.

The NCSM is based on solving for the nuclear many-body wave functions in a basis
of antisymmetrized products (Slater determinants) of single-nucleon wave
functions, which are usually taken as harmonic oscillator orbitals.  Written in
terms of this basis, the Schr\"odinger equation becomes a matrix eigenproblem.
However, calculations can only be done with a finite basis, and the accuracy of
results depends on how well the true solution to the Schr\"odinger equation for
the many-body wave function can be approximated in this truncated basis.

In practice, the NCSM basis is truncated by keeping only Slater determinants in
which the nucleons have at most some maximum number $\Nmax$ of oscillator
excitations.  The numerical accuracy of the solution can, in principle, be made
arbitrarily good by increasing $\Nmax$, but the number of basis states, and thus
the dimension of the matrix eigenproblem, grows rapidly with $\Nmax$
(\textit{e.g.}, reaching $\sim2.5\times 10^8$ for the largest calculations for
$\isotope[7]{Be}$ and $\isotope[7]{Li}$, with $\Nmax=12$, shown here) and
eventually becomes prohibitive.  The accuracy of the calculation also depends
sensitively on the oscillator length scale (quoted here as an oscillator
frequency $\hw$) chosen for the basis.

To illustrate the convergence of NCSM results, let us momentarily restrict our
attention to one specific internucleon interaction, the Daejeon16
interaction~\cite{shirokov2016:nn-daejeon16}.  We carry out NCSM calculations
for this interaction, using the code
MFDn~\cite{maris2010:ncsm-mfdn-iccs10,aktulga2013:mfdn-scalability,shao2018:ncci-preconditioned},
to obtain energies and electromagnetic transition strengths, presented in
Fig.~\ref{fig:convergence}.  (Numerical tabulations of the calculated
observables in Fig.~\ref{fig:convergence} are provided in the Supplemental
Material~\cite{supplement}.)

For instance, for the ground state energy of $\isotope[7]{Be}$, we can see how
the values calculated in truncated bases approach the actual ground state energy
of this Schr\"odinger equation problem by examining
Fig.~\ref{fig:convergence}(a).  At fixed basis size, \textit{e.g.}, the
uppermost curve shows calculations for $\Nmax=2$, the calculated energy depends
upon $\hw$, but has a variational minimum at some value of $\hw$.  As the basis
is enlarged to $\Nmax=4$, $6$, \textit{etc.}, we obtain the successively lower
curves.  The approach to a converged result is indicated as the calculated
values become independent of $\Nmax$ (the curves lie atop one another) and
independent of $\hw$ (the curves become flat).

If we were considering $M1$ transitions, then convergence would readily be
obtained, as seen in Fig.~\ref{fig:convergence}(b) for the strength of the
lowest $M1$ transition.  We see that the prediction for $B(M1; 3/2^- \rightarrow
1/2^-)$ can be identified to well within \SI{0.1}{\nucmag^2}, even from
low-$\Nmax$ calculations.

However, the calculated $B(E2)$ values, shown in Fig.~\ref{fig:convergence}(c),
are still steadily changing as the basis size increases, even at $\Nmax=12$.
There is no clear indication from these truncated calculations as to what the
actual solution is for the $E2$ strength in the full, untruncated \textit{ab
initio} problem.  (There is perhaps at most a hint of a flattening of the curves
at the lower end of the $\hw$ range.)  The same general behavior holds for the
$B(E2)$ values calculated for $\isotope[7]{Be}$ (solid curves) and
$\isotope[7]{Li}$ (dashed curves), although with different overall scales.  In
fact, therein lies the essential observation~--- that the convergence of the two
$E2$ values follows a similar pattern, except for scale, and that their ratio
may therefore be stable.

For the ratio to be stable with respect to $\hw$ and $\Nmax$, the transition
strengths entering into the ratio must have the same overall form for their
convergence behavior, as functions of $\Nmax$ and $\hw$, differing only in an
overall normalization factor.  This is plausible if the wave functions of the
states involved have similar structure, but the convergence behavior of $E2$
observables is in general not well understood (see, \textit{e.g.},
Ref.~\cite{odell2016:ir-extrap-quadrupole} for a proposed functional form for
their convergence in the two-body system), and the degree of convergence of the
ratio is for now a matter to be determined empirically.

In NCSM calculations of $E2$ transitions within rotational bands in light
nuclei~\cite{caprio2013:berotor,maris2015:berotor2,caprio2015:berotor-ijmpe} it
has been found that, even though each of the $E2$ transition strengths within
the band is not individually converged, the \textit{ratios} of $E2$ strengths
within a rotational band already converge to approximately rotational ratios at
low $\Nmax$ (see Fig.~8 of Ref.~\cite{caprio2015:berotor-ijmpe}).  We now
similarly consider a ratio of $E2$ strengths across analog transitions in mirror
nuclei, in Fig.~\ref{fig:convergence}(d).  Given that the initial and final
states in the $\isotope[7]{Be}$ $3/2^-\rightarrow 1/2^-$ transition are isobaric
analog states to those in the $\isotope[7]{Li}$ $3/2^-\rightarrow 1/2^-$
transition, it is not unreasonable that we find similar convergence properties
for their wave functions, and thus transition observables, in NCSM calculations.
Indeed, it is seen that the calculated ratio of $E2$ strengths in
$\isotope[7]{Be}$ and $\isotope[7]{Li}$ converges rapidly, by $\Nmax\sim6$, to a
value in the range $\sim3.0$--$3.1$.

The \textit{ab initio} $E2$ ratio predictions in Fig.~\ref{fig:convergence}(d)
are based on one particular choice of internucleon interaction, and we must
understand the sensitivity of these predictions to the input interaction.  Of
course, without converged predictions of $E2$ observables in NCSM calculations,
it has not been possible to study the sensitivity, to the choice of internucleon
interaction, of the predictions for \textit{absolute} $E2$ strengths. However,
the $\isotope[7]{Be}$/$\isotope[7]{Li}$ $B(E2)$ ratio provides a common ground
for comparison across different internucleon interactions.

\begin{figure}[t]
  \centering
  \includegraphics[width=\hsize]{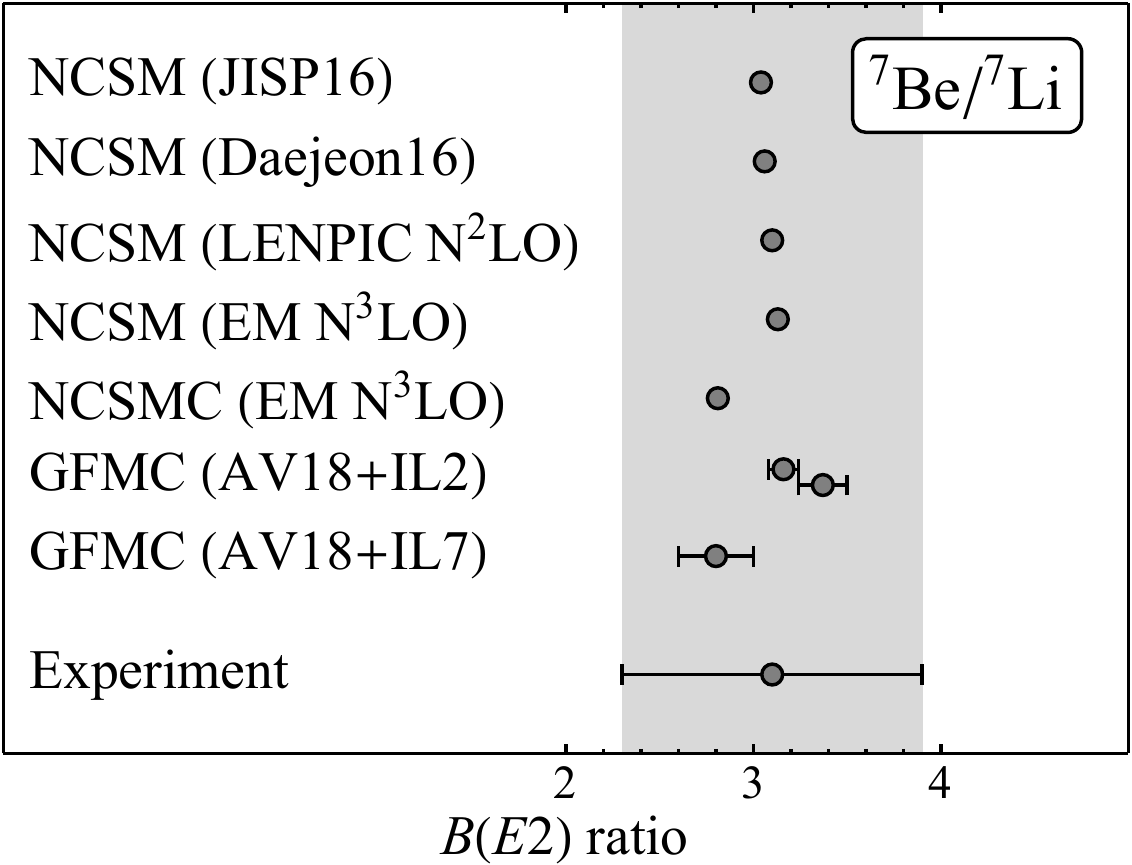}
  \caption{\textit{Ab initio} predictions for the ratio of the $\BEtwo$ strength
    in $\isotope[7]{Be}$ to that in $\isotope[7]{Li}$, obtained by various
    many-body solution methods and for various internucleon interactions (see
    text).  The experimental ratio is shown for comparison with $1\sigma$
    uncertainties.  Further details of the calculations may be found in the text
    (see also footnote~\ref{fn-basis-details}).}
  \label{fig:be2-ratio}
\end{figure}

\begin{table}
  \caption{ \textit{Ab initio} GFMC predictions for absolute $\BEtwo$ strengths
    in $\isotope[7]{Li}$ and $\isotope[7]{Be}$.  Experimental values are shown
    for comparison. All values are given in \si{e^2\fm^4}.}
  \label{tab:be2}

  \begin{tabular}{lllll}
    \hline\hline
    Method                         & Interaction & $\isotope[7]{Li}$ & $\isotope[7]{Be}$      & Reference                                                     \\
    \hline
    GFMC                           & AV18+IL2    & 8.09(17)          & 25.6(3)                & \cite{pervin2007:qmc-matrix-elements-a6-7}\textsuperscript{a} \\
    {}                             & AV18+IL2    & 8.15(20)          & 27.5(8)                & \cite{pervin2007:qmc-matrix-elements-a6-7}\textsuperscript{b} \\
    {}                             & AV18+IL7    & 7.81(45)          & 22.2(11)               & \cite{Pastore2013}                                            \\
    \hline
    \multicolumn{2}{l}{Experiment} & 8.3(5)      &                   & \cite{npa2002:005-007}                                                                 \\
    {}                             &             &                   & 26(6)(3)               & Present                                                       \\
    \hline\hline
  \end{tabular}
  \\
  \scriptsize
  \textsuperscript{a} Computed using ``Type I'' trial wave functions and reprojected interactions (AV8$'$+IL2$'$).\\
  \textsuperscript{b} Computed using ``Type II'' trial wave functions and reprojected interactions (AV8$'$+IL2$'$).
\end{table}

Predictions for the $B(E2)$ ratio from NCSM calculations based on different
interactions are compared in Fig.~\ref{fig:be2-ratio}.%
\footnote{\label{fn-basis-details}
  For reference, we detail the basis parameters for the calculations yielding
  the results in Fig.~\ref{fig:be2-ratio} and record the numerical values for
  the ratios plotted in this figure: For the JISP16, Daejeon16, and LENPIC
  interactions, the basis parameter $\hw$ is chosen at the approximate
  variational minimum for the ground state energy.  The ratios obtained in these
  NCSM calculations are $3.04$ for JISP16 ($\Nmax=16$, $\hw=\SI{20}{\MeV}$; see
  Tables II and IV of Ref.~\cite{Heng2017}), $3.06$ for Daejeon16 ($\Nmax=12$,
  $\hw=\SI{12.5}{\MeV}$), and $3.10$ for LENPIC \ntwolo{} ($\Nmax=12$,
  $\hw=\SI{27.5}{\MeV}$).  For EM \nthreelo{}, the ratios are based on $E2$
  strengths extracted in Ref.~\cite{QuaglioniPC} from the wave functions
  computed in Ref.~\cite{Dohet-Eraly2016}.  These yield a ratio of $3.13$ from
  the NCSM calculations ($\Nmax=10$, $\hw=\SI{20}{\MeV}$) and $2.81$ from the
  NCSMC calculations (which combine this NCSM basis for the $A=7$ system with an
  $\Nmax=12$ RGM cluster basis).  The GFMC ratios are obtained from the $B(E2)$
  values already described above in Table~\ref{tab:be2}, while the uncertainties
  shown on the ratios are obtained by combining the statistical uncertainties on
  the individual calculated $B(E2)$ values in quadruature: the resulting ratios
  are $3.16(8)$, $3.37(13)$, and $2.8(2)$, corresponding to the first three rows
  in Table~\ref{tab:be2}, respectively.
}
In addition to the Daejeon16 calculations already discussed, we carry out NCSM
calculations with the LENPIC \ntwolo{} chiral EFT
interaction~\cite{binder2016:lenpic-chiral}. (Numerical tabulations of the
calculated observables as functions of $\Nmax$ and $\hw$ are provided in the
Supplemental Material~\cite{supplement}.) We also compare with ratios extracted
from previous NCSM calculations for the JISP16~\cite{shirokov2007:nn-jisp16}
interaction, taken from Ref.~\cite{Heng2017}, and  the classic Entem-Machleidt
(EM) \nthreelo{} chiral EFT interaction~\cite{entem2003:chiral-nn-potl}, taken
from Refs.~\cite{Dohet-Eraly2016,QuaglioniPC}.

The Daejeon16 interaction, which we have considered so far above, is obtained
from the two-body part of the classic Entem-Machleidt (EM) \nthreelo{} chiral
EFT interaction with \SI{500}{\MeV} ultraviolet
regulator~\cite{entem2003:chiral-nn-potl}, which is then softened via a
similarity renormalization group (SRG) transformation and adjusted via a
phase-shift equivalent transformation to describe light nuclei, as detailed in
Ref.~\cite{shirokov2016:nn-daejeon16}. The JISP16~\cite{shirokov2007:nn-jisp16}
interaction, in contrast, is derived from nucleon-nucleon scattering data by
$J$-matrix inverse scattering, yielding a two-body interaction which is likewise
adjusted via a phase-shift equivalent transformation to describe light nuclei.
As an example of a modern chiral EFT interaction, we use the two-body component
of the recently-developed LENPIC \ntwolo{} interaction with semi-local
coordinate space regulator ($R=\SI{1}{\fm}$)~\cite{binder2016:lenpic-chiral}.
The ratios shown for the EM \nthreelo{} interaction are based on the two-body
component of this interaction~\cite{entem2003:chiral-nn-potl}, softened via a
similarity renormalization group (SRG) transformation to a resolution scale of
$\Lambda=\SI{2.15}{\fm^{-1}}$.

The notable point in Fig.~\ref{fig:be2-ratio} is the remarkable consistency of
the predictions for the $B(E2)$ ratio from the \textit{ab initio} NCSM
calculations, essentially independent of the choice of interaction.  We may
compare these with an experimental ratio of $3.1(8)$, obtained based on the
experimental values in Table~\ref{tab:be2} (the uncertainties on the
$\isotope[7]{Be}$ and $\isotope[7]{Li}$ $E2$ strengths have simply been treated
as uncorrelated and combined in quadrature).  The \textit{ab initio} NCSM
predictions for the ratio are all within the experimental uncertainty range and
agree with the measured result.

The $B(E2)$ ratio furthermore provides a means of comparing predictions, not
just across interactions, but across different many-body solution methods, as
also shown in Fig.~\ref{fig:be2-ratio}.  In such a comparison, we should keep in
mind that convergence behaviors differ across many-body methods, so the
convergence of the calculated ratio must ultimately be reassessed for each
method.

The $E2$ ratios obtained using the different sets of GFMC calculations from
Table~\ref{tab:be2} scatter signficiantly more than the ratios obtained from the
NCSM calculations, as shown in Fig.~\ref{fig:be2-ratio}.  However, they are
approximately consistent with the NCSM values to within the statistical
uncertainties.

By explicitly including cluster degrees of freedom into the NCSM basis, the
no-core shell model with continuum (NCSMC)
approach~\cite{romeroredondo2016:6he-correlations} attempts to attain more
rapidly convergent calculations.  NCSMC calculations for $\isotope[7]{Li}$ and
$\isotope[7]{Be}$ are presented in Ref.~\cite{Dohet-Eraly2016}, for the EM
\nthreelo{} interaction.  These combine an $A=7$ ($\isotope[7]{Li}$ or
$\isotope[7]{Be}$) NCSM basis at $\Nmax=10$ and $\hw=\SI{20}{\MeV}$ with
microscopic cluster states (involving $\isotope[4]{He}$ and $\isotope[3]{H}$ or
$\isotope[3]{He}$ clusters) at $\Nmax=12$.  The $B(E2)$ values obtained from
these NCSMC wave functions are $\SI{7.12}{e^2\fm^4}$ for $\isotope[7]{Li}$ and
$\SI{20.02}{e^2\fm^4}$ for $\isotope[7]{Be}$~\cite{QuaglioniPC}, comparable to
experiment (the corresponding ratio is shown in Fig.~\ref{fig:be2-ratio}).  For
comparison, NCSM calculations with the same interaction yield $B(E2)$ strengths
ranging from $\SI{2.66}{e^2\fm^4}$ at $\Nmax=6$ to $\SI{3.486}{e^2\fm^4}$ at
$\Nmax=10$ for $\isotope[7]{Li}$, or $\SI{8.45}{e^2\fm^4}$ at $\Nmax=6$ to
$\SI{10.901}{e^2\fm^4}$ at $\Nmax=10$ for $\isotope[7]{Be}$~\cite{QuaglioniPC}.

In summary, \textit{ab initio} predictions obtained using a variety of realistic
internucleon interactions and different many-body solution methods give
remarkably robust and consistent predictions for the $B(E2)$ ratio between the
mirror transitions in $\isotope[7]{Be}$ and $\isotope[7]{Li}$
(Fig.~\ref{fig:be2-ratio}), with a spread of only $\sim2\%$ in the NCSM results,
or $\lesssim20\%$ if the GFMC and NCSMC calculations are considered as well. The
current experimental results (Table~\ref{tab:be2}) for the strengths in
$\isotope[7]{Be}$ and $\isotope[7]{Li}$ are consistent with the calculated NCSM
GFMC results, within the one-sigma uncertainty range on the experimental value.
It should be noted that the experimental value we have used for the $B(E2)$
strength in $\isotope[7]{Li}$ is the evaluated value~\cite{npa2002:005-007}, but
conflicting results may be found among the various Coulomb excitation
measurements and
analyses~\cite{Hausser1972,Bamberger1972,Hausser1973,Vermeer1984c,Vermeer1984b,Weller1985,Barker1989,Vermeer1989,Voelk1991}.
More precise experimental values, for both mirror isotopes, would provide a more
stringent test of the \textit{ab initio} theory.

\section{Summary and Outlook}

We have performed a radioactive beam Coulomb excitation experiment to measure
the $\BEtwo$ transition strength in \isotope[7]{Be} for the first time, with the
aim of testing the ability of \textit{ab initio} theory to provide accurate
predictions of electromagnetic observables. Although $E2$ observables can
present a computational challenge to the \textit{ab initio} many-body solution
methods, due to their sensitivity to the long-range components of the wave
functions, we have found that the ratios of $E2$ strengths for isospin mirror
transitions are robustly converged in NCSM calculations. The calculated ratios
are remarkably consistent across internucleon interactions and many-body
solution methods.

We combine our measured $\isotope[7]{Be}$ transition strength
[$B(E2;3/2^-\rightarrow1/2^-)=\SI{26(6)(3)}{e^2\fm^4}$] with the known
$\isotope[7]{Li}$ transition strength to obtain an experimental ratio of
$B(E2)_{\isotope[7]{Be}}/B(E2)_{\isotope[7]{Li}}=\num{3.1(8)}$. This is
generally consistent, within uncertainty, with the \textit{ab initio}
predictions, which cluster around $\sim3.1$, well within the experimental
one-sigma uncertainty range.

To provide a more comprehensive set of precision electromagnetic tests of
\textit{ab initio} theory for light nuclei, the $B(E2)$ ratio should be
investigated for additional mirror transitions (and possibly nonmirror
transitions), such as in the $A=8$ isobars $\isotope[8]{Li}$ and
$\isotope[8]{B}$.  A previous Coulomb excitation measurement has yielded the
$E2$ transition strength in $\isotope[8]{Li}$ [$B(E2; 2^+ \rightarrow 1^+) =
55(11)$ e$^2$fm$^4$] \cite{Brown1991}, but the $E2$ transitions strength in
$\isotope[8]{B}$ is currently unknown.  While so far only unconverged NCSM
calculations of the $\isotope[8]{Li}$ $B(E2; 2^+ \rightarrow 1^+)$ transition
strength have been discussed~\cite{Maris2013}, we expect that the convergence
limitations in the NCSM calculations can again be overcome by considering the
ratio with the $\isotope[8]{B}$ mirror transition strength.

New data on electromagnetic observables for these and other light nuclei would
give tighter constraints on the various \textit{ab initio} descriptions that are
now available and either validate or challenge our understanding of the
microscopic origins of nuclear structure in this region. Such tests of nuclear
theory will both validate and contribute to the development of a higher degree
of predictive power for \textit{ab initio} approaches. These approaches promise
to have significant implications not only for nuclear structure, but for nuclear
interactions and nuclear astrophysics as well, such as in the calculation of
low-energy $S$ factors~\cite{Nollett2001,Neff2011,Dohet-Eraly2016}.

\section*{Acknowledgments}
We thank S.~Quaglioni and collaborators for sharing their NCSMC results for the
$B(E2; 3/2^- \rightarrow 1/2^-)$ for $\isotope[7]{Li}$ and $\isotope[7]{Be}$,
P.~Navratil and S.~Quaglioni for comments on the manuscript, and K.~Nollett for
pointing us towards the results of S.~Pastore. We thank A.~Moro for performing
coupled-channels calculations for our experiment. We also thank X. Li and J.
Riggins for their help during data collection. Additionally, we acknowledge the
Clovershare collaboration for the use of the HPGe Clover detectors at the NSL at
the University of Notre Dame. This work was supported by the U.S.\ National
Science Foundation under Grants No. PHY 17-13857, No. PHY 14-01343, and No. PHY
14-30152 and the U.S.\ Department of Energy under Grant No. DE-FG02-95ER-40934.
TRIUMF receives federal funding via a contribution agreement with the National
Research Council of Canada.  This research used computational resources of the
National Energy Research Scientific Computing Center (NERSC), which is a DOE
Office of Science User Facility (Contract~DE-AC02-05CH11231).

\bibliographystyle{apsrev4-1}
\bibliography{bib-mac/master,Clovershare,bib-mac/mc,bib-mac/theory,bib-mac/data,bib-mac/books}

\end{document}